\documentclass[12pt,preprint]{aastex}
\begin{document}
\title{A Magnetic Flux Tube Oscillation Model for QPOs in SGR Giant Flares}
\author{Bo Ma, Xiang-Dong Li, and P. F. Chen}
\affil{Department of Astronomy, Nanjing University, Nanjing 210093,
China;\\ xiaomabo@gmail.com, lixd@nju.edu.cn, chenpf@nju.edu.cn}

\begin{abstract}
Giant flares from soft gamma-ray repeaters (SGRs) are one of the most
violent phenomena in neutron stars. Quasi-periodic oscillations
(QPOs) with frequencies ranging from 18 to 1840 Hz have been
discovered in the tails of giant flares from two SGRs, and were ascribed
to be seismic vibrations or torsional oscillations of magnetars.
Here we propose an alternative explanation for the QPOs in terms of
standing sausage mode oscillations of flux tubes in the magnetar
coronae. We show that most of the QPOs observed in SGR giant flares
could be well accounted for except for those with very high
frequencies (625 and 1840 Hz). 
\end{abstract}

\section{Introduction}
Soft gamma-ray repeaters (SGRs) are neutron stars that exhibit
sporadic burst activities most prominently in soft gamma-rays
(Norris et al. 1991). Besides normal bursts with energy $E\sim
10^{41}$ ergs, enormously energetic giant flares (with $E\sim
10^{44}-10^{46}$ ergs) have also been observed, for example, from
SGR 0526$-$66 in 1979 (Mazets et al. 1979), from SGR 1900$+$14 in
1998 (Hurley et al. 1999; Feroci et al. 1999) and from SGR 1806$-$20
in 2004 (Hurley et al. 2005; Terasawa et al. 2005; Palmer et al.
2005). The giant flares generally start with an initially rising
spike lasting $\sim 1$ s and then evolve into a decaying phase which
lasts hundreds of seconds. Theoretically, SGRs are thought to be
magnetars, neutron stars with surface magnetic fields strengths of
$\sim 10^{14}-10^{15}$ G (Thompson \& Duncan, 1993; Duncan \&
Thompson, 1994; Thompson \& Duncan, 1996, 2001), or neutron stars
with normal magnetic fields accreting from a disk formed from the
fallback in supernova explosions (e.g., Alpar 2001).

During the three SGR giant flares mentioned above, quasi-periodic
oscillations (QPOs) were identified both in the initially rising
spike and in the decaying tail (Barat et al. 1983; Terasawa et al.
2005; Palmer et al. 2005; Israel et al. 2005; Strohmayer \& Watts
2005; Watts \& Strohmayer 2006; Terasawa et al. 2006). These QPOs
seem to provide independent evidence for superstrong magnetic fields
in SGRs (Vietri, Stella, \& Israel 2007). In the most popular
models, they are explained as global seismic vibration modes of
magnetars (Hansen \& Cioffi 1980; Schumaker \& Thorne 1983;
McDermott, van Horn \& Hansen 1988; Strohmayer 1991; Duncan 1998).
Other explanations have also been proposed. Levin (2006) argued that
the QPOs may be driven by the global mode of the MHD fluid core of
the neutron star and its crust, rather than the mechanical mode of
the crust. Following this idea, Sotani et al. (2007) recently made
two-dimensional numerical simulations, and found two families of
torsional Alfv\'{e}n oscillations which may explain some of the
observed QPOs. Coupling neutron star's elastic crust and fluid core,
Glampedakis et al. (2006) used a simple toy model to provide
explanations for some of the QPOs observed. Besides, magnetospheric
current variation induced by the crust oscillation was proposed for
the modulation of X-ray flux by Timokhin, Eichler, \& Lyubarsky
(2007), and they showed that radial oscillation with an amplitude of
one-hundredth of the neutron star radius can account for the
observed QPO flux fraction.

In this paper we propose an alternative explanation for QPOs
observed in the SGR giant flares. We assume that part of the plasma
ejected during the giant flares are trapped by the magnetic fields,
and then form magnetic flux tube structures, similar as in the solar
corona. Oscillations of  magnetic loops or tubes in the solar corona
have been observed by a lot of authors (see Aschwanden et al. 1999
for a review). We argue that tube oscillations in SGR magnetospheres
may give rise to some of the QPOs observed during the giant flares.
We discuss the plausible tube oscillation modes in the SGR
magnetosphere in \S 2, and compare the possible oscillation
frequencies with observations based on the fireball model of
Thompson \& Duncan (1995) in \S 3. The possible implications of our
model are discussed in \S 4.

\section{Magnetic flux tube oscillations}

In solar physics QPOs with periods of several minutes in coronal
loops have been detected and successfully interpreted in terms of
MHD waves (Aschwanden et al. 1999; Nakariakov et al. 1999).
Following this idea, we suppose that these MHD waves could also be
excited in the magnetar coronae when starquakes of a magnetar shear
its external magnetic field, which becomes nonpotential and threaded
by an electric current (Beloborotov \& Thompson 2007). As pointed
out by Robets, Edwin \& Benz (1984), MHD tube oscillations in the
sausage mode can modulate the cross section of the tube, and hence
its density and radiation flux, while oscillations in the kink mode
do not change the loop density in the first order and thus cannot
modulate the plasma radiation (Aschwanden et al. 1999). So we rule
out the kink mode oscillations for the QPOs observed in SGRs.
Furthermore, observations of the QPOs show that they can last
hundred to thousand cycles of the oscillation periods (Israel 2005;
Terasawa et al. 2005; Palmer et al. 2005; Strohmayer \& Watts 2005;
Watts \& Strohmayer 2006). Accordingly, we exclude the propagating
sausage mode oscillations because of their rather short damping
timescale (about  tens of the oscillation period), and consider only
the standing sausage mode oscillations.

The flux tube can be roughly described as a cylinder of radius $a$
and length $L$. The magnetic field strength, temperature, mass and
number densities are denoted as $B_0$, $T_0$, $\rho_0$, $n_{0}$
inside the tube, and $B_{\rm e}$, $T_{\rm e}$, $\rho_{\rm e}$,
$n_{\rm e}$ outside the tube, respectively. Magnetoacoustic
oscillations of a magnetic tube have been studied thoroughly, and the derived
oscillation periods $\tau_{\rm fast}$ for the fast standing sausage mode
(Wentzel 1979; Edwin \& Roberts 1983; Roberts et al. 1984), and $\tau_{\rm slow}$ for the slow standing
sausage mode (Edwin \& Roberts 1983; Roberts et al. 1984) are expressed as follows:
\begin{equation}
\tau_{\rm fast}=\frac{2{\pi}a}{c_{\rm k}}=4\pi^{3/2}a(\frac{\rho_0+
\rho_{\rm e}}{B_0^2+B_{\rm e}^2})^{1/2} \simeq
2.0\times10^{-5}\frac{a_{6}n_{28}^{1/2}}{{B_{14}}}\,\rm s,
\end{equation}
\begin{equation}
\tau_{\rm slow}=\frac{2L}{jc_{\rm t}} =\frac{2L}{jc_{\rm
s}}[1+(\frac{c_{\rm s}}{v_{\rm A}})^{2}]^{1/2}\simeq
0.015\frac{L_{6}}{jT_{10}^{1/2}}\,\rm s,
\end{equation}
where $c_{\rm k}$, $c_{\rm t}$, $c_{\rm s}$, $v_{\rm A}$ are the
kink speed, slow magnetoacoustic speed, sound speed,  and Alfv\'en
speed, respectively (Roberts 2000), $L_{6}=L/10^{6}$ cm,
$a_6=a/10^6$ cm, $n_{28}=n/10^{28}$ cm$^{-3}$, $B_{14}=B/10^{14}$ G,
and $T_{10}=kT/10$ kev, $j$ is the number of nods and $j=1$ for the
fundamental mode. In deriving Eqs.~(1) and (2), we have made the
approximations that (i) the sound speed in the magnetar
magnetosphere is much smaller than the Alfv\'en speeds, i.e.,
$c_{\rm s} \ll v_{\rm A}$, so the slow magnetoacoustic speed $c_{\rm
t}$ is close to the sound speed $c_0$ inside the flux tube; (ii) the
magnetic field strength inside a magnetic flux tube $B_0$ is
comparable to that in the environment $B_{\rm e}$, whereas the
plasma density $\rho_0$ inside the tube is much higher than
$\rho_{\rm e}$ in its surroundings. Thus, we have $c_{\rm k}\simeq
\sqrt{2}v_{\rm A}$. We also ignore the effects of gravity on the
flux tube because of the super-strong magnetic field of magnetars.
With the typical values for the parameters of the magnetar tube, the
above equations indicate that the oscillation frequencies in the
fast standing sausage mode are too hight to be compatible with the
observed QPOs in SGRs, and we are left with only the slow standing
sausage mode oscillations.

\section{QPOs in giant flares}
In this section, we investigate how well the slow sausage mode
oscillation frequencies can match the QPOs frequencies observed in
SGR giant flares. The QPOs observed in the initial spike phase and
in the flare tail phase are discussed separately.

\subsection{QPOs in the initial flare spike phase ($t<1$ s)}
Geotail spacecraft mission was originally aimed to study the
structure and dynamics of the tail region of the magnetosphere of
the Earth. With it 50 Hz and 48 Hz QPOs were detected at $t=45-175$
and $430-567$ ms after the onset of the giant flare in SGR 1806$-$20
in 2004 (Terasawa et al. 2006). The oscillation periods ($\tau\sim
20$ ms) are similar to the period $\tau\sim 23$ ms of the QPOs from
SGR 0526$-$66's giant flare in 1979 (Barat et al. 1983), but no
similar phenomena have been seen in the SGR 1900$+$14 flare.

The physical picture in our flux tube oscillation scenario is as
follows: after the onset of the giant flare, hot plasma was ejected
into the magnetosphere above the surface of the SGR. Because of the
super-strong field strength of the magnetar, the plasma could only
move along the field lines, from one to another footpoint of each
field line on the star's crust, and could not get out of the
confined structure of the field. This is why a fireball is formed.
We assume that the magnetic tubelike structure(s) were subject to
various types of oscillations excited by the turbulence at the
footpoints. For the slow sausage mode oscillation, the period is
related to the initial e-folding rising time $t_{\rm rise}=L/c_{\rm
s}$, which is 9.4 ms\footnote{Schwartz (2005) measured the e-folding
rising time to be $\sim 4.9$ ms from the observations on giant flare
from SGR 1860$-$20 with Chinese Double Star polar spacecraft.} in
the giant flare from SGR 1806$-$20 according to Geotail spacecraft
observations (Tanaka et al. 2007), i.e.,
\begin{equation}
\tau_{\rm slow}=\frac{2L}{jc_{\rm t}}\simeq \frac{2L}{jc_{\rm
s}}\sim 2t_{\rm rise}\sim 18.8\,{\rm ms}
\end{equation}
for $j=1$ (we do not need to consider the gravitational redshift
effect in $\tau_{\rm slow}$ here because the observed rising time
$t_{\rm rise}$ has already included the redshift factor). This
period is very close to the observed value $\sim 20$ ms.

Unfortunately, there was no QPO detected in the initial rising phase
of the giant flare in SGR 1900$+$14 with rising time $t_{\rm
rise}\sim3.1$ ms, and no rising time was measured in the 1979 March
giant flare event with an initial $\sim 43$ Hz QPO. So in the three
giant flares ever detected we have only one event on SGR 1806$-$20
to examine our explanation for the QPOs in the initial spike phase,
which should be testified by future detailed monitoring of SGRs'
giant flares.

\subsection{QPOs in the flare tail phase ($t>20$ s)}
\begin{figure}
\plotone{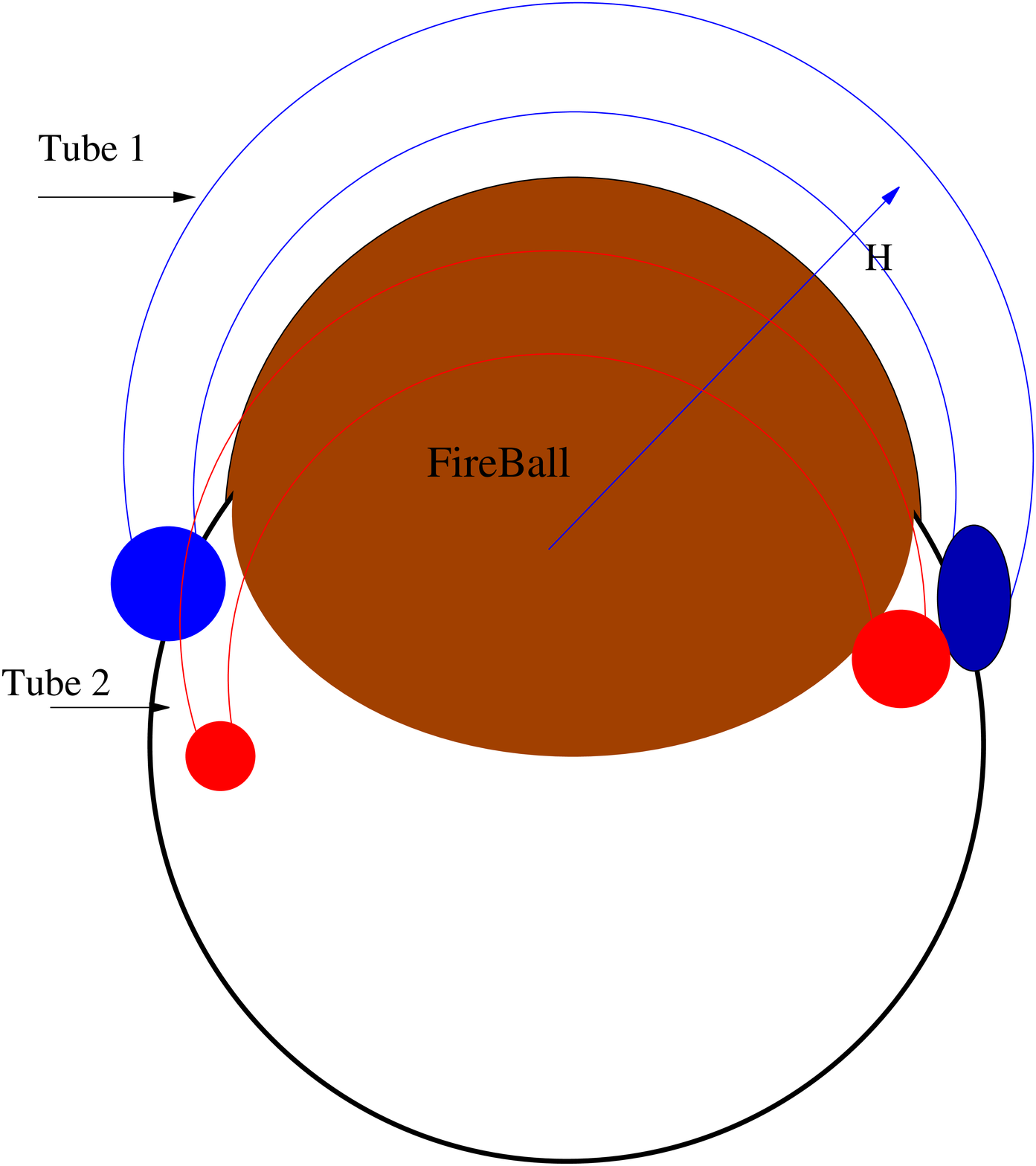}
\epsscale{.50}
\caption{Sketch of the flux tube formed in the tail of SGRs giant
flare. The height of the tube $H$
has also been indicated here.
\label{tube}}
\end{figure}

In the tail phase of the giant flare, declining of the radiation
flux is explained as the shrinking of the fireball surface area
(Thompson \& Duncan 1995). After the fireball has evaporated to
somewhat a smaller size, the plasma left from the fireball may form
flux tubes above the fireball, and the slow sausage oscillations of
such tubes may cause the QPO phenomena. The pulse phase-dependence
of the QPO amplitude indicate that these oscillations are intrinsic
to the neutron star surface (Israel et al. 2005; Strohmayer \& Watts
2006). This could be naturally explained as that the magnetic flux
tube's footpoints are anchored at certain regions of the neutron
star's crust. Sketche of the structure of such flux tubes is shown
in Fig.~\ref{tube}.
The length of the tube $L$ is related to its height
$H$ as
\begin{equation}
L\simeq b\pi H
\end{equation}
with the geometry index $b\sim1$. Combine Eqs.~(2) and Eqs.~(4), we
have the oscillation frequency in the slow standing sausage mode as
\begin{equation}
f_{\rm slow,j}\simeq 21[1-\frac{2GM_{\rm NS}}{(R_{\rm NS}+H)c^2}]^{1/2}\frac{jT_{\rm 10}^{1/2}}{bH_6}\,\rm Hz,
\end{equation}
where $H_6= H/10^6$ cm, and the term in the bracket is attributed to
the gravitational redshift effect ($M_{\rm NS}$ and $R_{\rm NS}$ are
neutron star mass and radius, respectively).

We then choose typical values of $T$ and $H$ in the tail phase of
the giant flare to estimate the QPO frequencies. In Feroci et al.
(2001), a blackbody component with $T=9.3$ keV, which accounts for
$\sim 85\%$ of the total energy released above $25$ keV, was derived
at $t=65-195$ s after the onset of the SGR 1900$+$14 giant flare. So
we may set $kT\simeq 10$ keV for the flare in the pulsating tail
phase (Thompson \& Duncan 1995; Thompson \& Duncan 2001; Hurley et
al. 2005; Boggs et al. 2007). Previous studies of giant flares
suggested a typical length scale $L\sim 10$ km for the `fireball'
formed in the burst (Thompson \& Duncan 1995; Thompson \& Duncan
2001; Hurley et al. 2005; Boggs et al. 2007), so we may take
$H_6\simeq 1$ for the flux tube. These values give $f_{\rm slow}\sim
19$ Hz for the fundamental mode ($j=1$) in a neutron star of mass
$M_{\rm NS}=1.4\,M_\odot$ and radius $R_{\rm NS}=10$ km. Consider
its harmonic oscillations, most of the observed QPOs (summarized in
Table 1) may be well explained, except those with very high
frequencies of $625$ Hz and $1840$ Hz, because excitations of
oscillations with $j>20$ must be very difficult.

\subsection{Modulation of the Radiation}
In last section we showed that the slow sausage mode oscillations of
flux tubes seem to be consistent with the QPOs observed. Now we move
to the question how such tube oscillations modulate the radiation
observed in the tail phase of the giant flare. The mass of the
plasma in the flux tube can be estimated to be $\triangle M \sim
10^{23}$ g from Eq.~(22) in Thompson and Duncan (1995), assume the
total giant flare energy $E\sim 3\times10^{46}$ ergs (Cameron et al.
2005). The plasma density is then $\rho=\triangle M/\triangle V\sim
10^{23}/10^{18}\sim 10^5$ g cm$^{-3}$, where $\triangle V$ is the
tube volume. The Rosseland mean scattering cross-section in the
direction parallel to the magnetic field is (Thompson \& Duncan
1995)
\begin{equation}
\sigma_{\rm es}=2.2\times10^9T^2B^{-2}\sigma_{\rm
T}\simeq1.5\times10^{-15}T^2B^{-2},
\end{equation}
where $\sigma_{\rm T}=(8\pi/3)(e^2/m_{\rm e}c^2)^2$ is the Thomson
scattering cross-section. Then the optical depth of the tube is
\begin{equation}
\tau=n\sigma_{\rm es}L=\frac{\rho}{m_{\rm p}}\sigma_{\rm es}
L\sim10^9T_{\rm 10 kev}^2B_{14}^{-2}L_{6},
\end{equation}
suggesting that the flux tube is optically thick. Since the fireball
itself is also optically thick (Thompson \& Duncan 1995), the
fraction of the radiation flux from tube in the total flux from the
fireball can be simply expressed as the ratio of their surface area 
perpendicular to the line of sight
\begin{equation}
\frac{F_{\rm{tube}}}{F_{\rm{fireball}}}\sim
\frac{S_{\rm{tube}}}{S_{\rm{fireball}}} \sim \frac{2 aL}{\pi
r_{\rm{fireball}}^2}\sim \frac{2a}{r_{\rm{fireball}}},
\end{equation}
if we assume that the thermal temperatures of the fireball and the
tube are roughly the same (about $10$ keV).  Here $a$ is the radius 
of the flux tube, $r_{\rm fireball}$ is the radius of the fireball, 
and $L\sim \pi r_{\rm fireball}$. As the flux tube oscillates, its 
cross-section and surface area vary, and so does the thermal emission 
from the tube. The amplitude of the QPOs can be derived to be
\begin{equation}
\frac{\delta F_{\rm{tube}}}{F_{\rm{fireball}}}\sim\frac{\delta a}
{a}\frac{ S_{\rm{tube}}}{ S_{\rm{fireball}}}\sim\frac{\delta
a}{a}\frac{2a}{r_{\rm{fireball}}}.
\end{equation}
This amplitude is consistent with the observational values $\sim
10\%-20\%$ (Israel et al. 2005; Strohmayer \& Watts 2006; Watts \&
Strohmayer 2006), only when the radial flux tube oscillation
amplitude $\delta a$ is in the range $\delta a/r_{\rm fireball}\sim5\%-10\%$. 
This may also explain why the QPOs emerged from the light curve after 
about $\sim 100$ s from the onset of the giant flare: at the very 
beginning of the giant flare, the fireball was so big that 
($r_{\rm fireball}$ is very large), the tube radiation was too 
weak compared with that from the fireball, to produce a detectable QPO.

It is noted that the QPOs are more likely to be detected in hard
X-ray band (Strohmayer \& Watts 2005; Watts \& Strohmayer 2006), and
their amplitudes in soft X-ray band are not as strong as in hard
X-rays. One example is the $84$ Hz QPO from SGR 1900+14 (Strohmayer
\&Watts 2005): Its amplitude increased from $<14\%$ in the $<18$ keV
band, to $(20 \pm 3)\%$ in the $12-90$ keV band, and $(26 \pm 4)\%$
in the $>30$ keV band. There are several possible reasons for this
QPO amplitude-energy band dependence. The first is the photon
splitting mechanism (Thompson \& Duncan 1995). Diffusion of photons
from the fireball is primarily in E-mode due to the different
scattering cross-sections between the two-polarization modes. Before
the E-mode photons reach the flux tube, the photons with energy
higher than $40$ keV still suffer serious photon splitting effect in
the magnetic fields higher than the quantum magnetic field (the QED
field $B_{\rm{QED}}=4.4\times10^{13}$ Gauss). This will produce an
excess in the $10-20$ kev in the spectrum from the fireball
(Lyubarsky 2002), decreasing the ratio of the photon fluxes from the
tube and from the fireball in this energy band and hence the QPO
amplitude. The second is that the cross section of electron
scattering decreases with decreasing photon energy, so the
low-energy photons seen by us come from deeper region in the
fireball, where temperature is higher (see Fig.~2 in Ulmer 1994).
This may also increase the low energy photon flux from the fireball
and reduce the amplitude of the QPOs in the same energy band. The
third is cyclotron scattering of thermal photons in the flux tube,
which occurs at energy (Ho \& Lai 2001; Zane et al 2001)
\begin{equation}
E= \hbar \frac{ZeB}{Am_{\rm p}c}=0.63(\frac{Z}{A})(\frac{B}{10^{14}
{\rm G}})\; \rm{kev},
\end{equation}
where $A$ and $Z$ are the mass number and charge number of the ion,
respectively. If we adopt $B\sim 10^{15}$ Gauss and $A=Z=1$ for
proton cyclotron resonance, this energy is $\sim 10$ keV. Such
proton resonant scattering will up-scatter the photons in soft X-ray
energy from the flux tube to higher energy (Lyutikov \& Gavriil
2006), thus increase the QPO amplitude in the hard X-ray band.

\subsection{Excitation of the QPOs}
In this subsection we derive the energy needed to excite the
oscillations of the flux tube, and the constraints on the possible
energy source. From Wang et al. (2003), the energy needed to excite
the slow sausage mode oscillations can can be estimated to be
\begin{equation}
\triangle E \sim  \frac{1}{2}\triangle M
v^{2}\sim\frac{1}{2}\triangle M (\frac{2\delta a c_{\rm
s}}{a})^{2}\sim\triangle MkT/m_{\rm p}\sim10^{39} \;\rm{ergs},
\end{equation}
where $\triangle M\sim10^{23}$ g and $kT=10$ keV are the mass and
temperature of the plasma within the flux tube.

According to recent investigations on the excitation of standing
slow mode oscillations in a flux tube (Taroyan et al. 2004, 2005),
the excitation energy should be injected into the tube within a
timescale similar to the oscillation period $\tau$ if the
oscillations are excited by the energy deposition at the footpoint
of the tube. If the energy deposition is through thermal conduction
(Spitzer 1962), the deposited energy would be $\triangle E \sim
F_{\rm c}S\tau$, with $F_{\rm c}\simeq
(1.84\times10^{-5}T^{5/2}\nabla T)/\ln\Lambda$ ergs$^{-1}$cm$^{-2}$
being the heating flux and $S$ the heating area. Taking $T\sim
10^{8}$ K, $\nabla T\sim 10^{2}$ Kcm$^{-1}$, $S\sim 10^{5}$
cm$^{2}$, and $\tau\sim 0.01$ s, we get $\triangle E \sim 10^{24}$
ergs, which is far less than that needed to excite the oscillations.
So we conclude that the excitation energy is unlikely to be
deposited through thermal conduction, but most likely by flare
activities at the footpoint(s) of the tube. This may explain why
there are no QPOs detected during the quiescent phase of SGRs.

\section{Discussion}
We have suggested the slow sausage oscillation modes of magnetic
flux tubes to explain most of the QPOs detected in SGRs during giant
flares. These QPOs generally last several rotational cycles, i.e.
tens of seconds, and present useful constraints on the damping
mechanisms for the QPOs. The damping of magnetic loop oscillations
in the solar corona has been extensively studied (Cally 1986;
Nakariakov et al. 1999; Ruderman \& Roberts 2002; Stenuit et al.
1999; Taroyan et al. 2004). Radial wave leakage and resonant
absorption are considered to be the main damping mechanisms. In our
case, we consider only the effect of resonant absorption, as
Ruderman \& Roberts (2002) pointed out that, if the mass density
inside the tube is great than that outside, the wave leakage effect
is not important. Considering the curvature of the tube, we can use
the damping time scale of kink mode oscillations as that of sausage
mode oscillation (Roberts 2000), which is given by (Ruderman \&
Roberts 2002)
\begin{equation}
\tau_{\rm
d}=\frac{2a}{\pi l}\frac{\rho_0+\rho_e}{\rho_0-\rho_e}\tau.
\end{equation}
Here we assume that the mass density varies in the annuls region
$a-l \leq r \leq a$ from $\rho_0$ to $\rho_{\rm e}$. From Eqs.~(18)
and (22) in Thompson \& Duncan (1995), we have
$a\sim2\gamma_b^{2}l\sim10^4l$. So $\tau_{\rm d}\sim10^4\tau$, which
is roughly consistent with the observational result $\tau_{\rm
d}\sim5\times10^3\tau$ for the 92 Hz QPO in SGR 1806$-$20 lasting
about 50 s. Note that the damping timescale is proportional to
oscillation period $\tau$, so higher frequency oscillations should
decay more quickly, which is also compatible with observations
(Strohmayer 2007). However, the above results are based on the
assumptions that the flux tube is thin, axis-symmetric with
homogeneous mass distribution, which may not be satisfied in the
real situation. Non-ideal effects, like the density stratification,
magnetic field curvature and twist, and thick tube limitation have
been studied in both theories and numerical simulations (Van
Doorsselaere et al. 2004a, b; Andries et al. 2005; Robert \& Viktor
2006; Arregui et al. 2007). These works show that the above effects
may only change the oscillation frequency by as much as $10\%-15\%$,
but may seriously damp the oscillation mode and reduce the damping
time scale. So the damping time scale with Eq.~(12) should be taken
as an upper limit

Observations indicate the QPOs frequencies may evolve with time. For
example, Israel et al. (2005) found a possible time evolution of the
QPO in  SGR 1806$-$20 with the frequency increasing from $92.5$ Hz
to $95$ Hz. From Eq.~(2) it is seen that the oscillation frequency
is determined by the length and temperature of the flux tube. Since
the flux tube's footpoints are anchored at the surface of the star
and confined by the super-strong magnetic fields that dominate the
plasma's motion, its structure and length may not change much. The
temperature of the flux tube is likely to remain nearly unchanged
during the life time of the QPOs (less than tens of seconds), if
there is no extra energy injected into the flux tube. So we would
not expect a considerable change in the QPO frequency. The QPO
amplitudes depend on the surface area of the fireball and the
oscillation amplitude of the flux tube (see Eq.~(9)), both of which
decrease with time during the tail phase of the giant flare. The
fact that  the QPO's lifetime ($\le 50$ s) is less than the
evaporation time of the fireball ($200-400$ s) implies that the tube
oscillations damp faster than the shrinking of the fireball. So in
this model the QPOs amplitude is predicted to decrease with time,
which could be testified in the future high time resolution
observations.

In the tube oscillation model the oscillation frequency is $f_{\rm
j}\propto j$, while in the seismic vibration model $f_{\rm
j}\propto[j(j+1)]^{1/2}$, where $j$ is the number of nodes
(McDermott, van Horn \& Hansen 1988 et al. 1988). To account for the
observed frequencies with seismic vibration, one has to use the
$j=2$, 4, and 6 modes but ignore the $j=3$ and 5 modes (Watts \&
Strohmayer 2006). Recently, Samuelsson \& Andersson (2007) have
derived a modified relation $f_{\rm j}\propto[(j-1)(j+2)]^{1/2}$ for
the torsional seismic modes using a general relativistic
formulation, but the problem why the $j=3$ and 5 modes do not exist
still remains. In this point of view, the tube oscillation model
seems to be more natural, since the $j=1$, 2, and 3 modes have all
been used.

There seem to exist three QPOs with different fundamental
frequencies (18 Hz, 26 Hz and 30 Hz) in SGR 1806$-$20 (Watts \&
Strohmayer 2006). In the seismic vibration model, only one
fundamental oscillation mode can exist (Israel 2005; Watts \&
Strohmayer 2006). However, in the tube oscillation model this could
be explained if there are three flux tubes oscillating in slightly
different fundamental modes. The existence of the multifrequencies
might be due to the deviations in the temperature, density, and
strength or topology of the magnetic field around the active regions
in the magnetosphere. It is noted, however, that it is difficult to
explain the $625$ Hz and $1840$ Hz QPOs observed in SGR 1900+14 in
the tube oscillation model. For these extremely high frequencies,
the $n=1$ and 3 torsional shear mode vibration (Piro 2005) might be
more reasonable explanation. This also suggests that the QPOs in the
giant flares may not be homogeneous, and may have different origins.

\begin{acknowledgements}
We are grateful to an anonymous referee for helpful comments that
helped improve the original manuscript. M.B. thank Yang Guo, Meng
Jin, and B. Roberts for their help and valuable suggestions. This
work was supported by Natural Science Foundation of China under
grant numbers 10573010 and 10221001.

\end{acknowledgements}

\clearpage

\clearpage
\begin{table}
\center
\begin{tabular}{c  c c}
\hline \hline
SGR 1806-20 & SGR 1900+14 & number of nodes\\
\hline \hline
18, 26, 30 & 28 & j=1 \\
  & 53 & j=2\\
92 & 84 &  j=3 \\
150  &  155 & j=5 \\
 625 &  & j=21(?) \\
 1840 &  & j=61(?) \\
\hline 
\end{tabular}
\label{datasum} \caption{QPO frequencies (in Hz) detected in the
tails of the giant flares from SGR 1900+14 and SGR 1806-20, together
with the number of nodes $j$ fitted in our flux tube oscillation
model (see text for more information). }
\end{table}

\end{document}